\documentclass[12pt]{iopart}

\usepackage{iopams}  
\usepackage{graphics}
\usepackage{epsfig}
\usepackage[latin9]{inputenc}
\begin{document}

\title{Some interesting features of new massive   gravity}

\author{Antonio Accioly$^{1,2}$, Jos\'{e} Helay\"{e}l-Neto$^1$, Eslley Scatena$^2$, Jefferson Morais$^1$, Rodrigo Turcati$^1$ and Bruno Pereira-Dias$^1$} 

\address{$^1$Laborat\'{o}rio de F\'{\i}sica Experimental (LAFEX), Centro Brasileiro de Pesquisas F\'{i}sicas (CBPF), Rua Dr. Xavier Sigaud 150, Urca, 22290-180  Rio de Janeiro, RJ, Brazil}
\address{$^2$Instituto de F\'{\i}sica Te\'{o}rica (IFT), S\~{a}o Paulo State University (UNESP), Rua Dr. Bento Teobaldo Ferraz 271, Bl. II-Barra Funda, 01140-070 S\~{a}o Paulo, SP, Brazil}

\eads{\mailto{accioly@cbpf.br}, \mailto{helayel@cbpf.br}, \mailto{scatena@ift.unesp.br}, \mailto{morais@cbpf.br},  \mailto{turcati@cbpf.br} and \mailto{bpdias@cbpf.br}}

\begin{abstract}
A proof that new massive  gravity  --- the massive 3D gravity model  proposed by Bergshoeff, Hohm and Townsend (BHT) --- is the only unitary system at the tree level that can be constructed by augmenting  planar gravity through the curvature-squared terms, is presented. Two interesting gravitational properties of the BHT model, namely, time dilation and time delay, which have no counterpart in the usual Einstein 3D gravity, are analyzed as well. 
\end{abstract}

\maketitle

\section{Introduction}
For too long physicists believed that gravity models containing fourth-  (or higher-) derivatives of the metric were doomed to failure by virtue of one detail: they entail unphysical ghost states of negative norm. The pure scalar curvature models, i.e., the fourth-order  gravity systems with Lagrangian $\mathcal{L} = R + \alpha R^2$ and which are  tree-level unitary, seemed to be the only exception to this rule.  Actually, these systems are conformally equivalent to Einstein gravity with a scalar field \cite{1}. Consequently, despite having fourth derivatives at the metric level, these models are ultimately second order in their scalar-tensor versions.  It is, therefore, perfectly understandable that just about two years ago the physical community were absolutely amazed to learn that  
a particular higher-derivative extension of  3D general relativity --- that is ghost-free at the  tree level ---  has been  found out by Bergshoeff, Hohm and Townsend (BHT) [2-14].  It was argued that this massive 3D gravity model, that is also known as ``new massive  gravity", is both unitary and power-counting   UV finite in its pure quadratic curvature limit \cite{15}, which, as it was pointed out by Ahmedov and Aliev \cite{10}, violates the standard paradigm of its ``cousins" in four dimensions \cite{16}.
New massive gravity is defined by the Lagrangian density

\begin{eqnarray}
\mathcal{L} = \sqrt{g}\left[ -\frac{2R}{\kappa^2} + \frac{2}{\kappa^2 m_2^2}\left(R_{\mu \nu}^2
- \frac{3}{8} R^2 \right) \right], 
\end{eqnarray}

\noindent where $\kappa^2 = 32\pi G$, with $G$ being the 3D analog of Newton's constant, and $m_2 \;(>0)$ is a mass parameter. It is worth noticing that the Lagrangian density given in (1) has a reversed Einstein-Hilbert (EH) term. On the other hand, a formal proof of the equivalence of the linearized version of the BHT model and the Einstein-Hilbert-Pauli-Fierz gravity was given in \cite{2}; incidentally, this proof was reviewed in \cite{17}. Nevertheless, the physical meaning of this equivalence is somewhat unclear; indeed, the linearized version of the BHT system is background diffeomorphism invariant, while the Pauli-Fierz theory is only invariant under the Killing symmetries of the spacetime (in particular, the 3D Minkowski space), which clearly shows that a better understanding of the symmetries is still lacking \cite{9}.  And what about the odd sign change of the EH term previously mentioned? At the linearized level,  Deser \cite{15} showed that the EH term breaks the Weyl invariance of the BHT model without the EH term and, consequently,  is responsible for giving mass to the graviton.  In other words, the higher-derivative terms provide the kinetic energy, whereas the EH term provides the  mass in this linearized model, thus explaining the weird sign change of the EH term. It is remarkable that the EH term gives origin to the mass in the linearized version of the BHT system by breaking the Weyl invariance and not the expected diffeomophism invariance \cite{9}. 

At this point it would be interesting to ask ourselves about the reason for doing research on massive gravitons. The increased interest in recent years in this subject is motivated, on the one hand, by the discovery of cosmic acceleration, which might be explained in terms of an infrared modification of general relativity that gives the graviton a small mass \cite{18}; on the other, by the conjecture that some theory involving massive gravitons could be the low energy limit of a noncritical string-theory underlying QED \cite{19}. As it is often done for so many other gravitational physical issues, it is advisable to consider first the possibilities for massive gravitons    in the simpler context of a 3D spacetime \cite{17}. The BHT model is accordingly  the ideal arena for such investigations.

Our aim in this paper is twofold.  
\begin{enumerate}
	\item To show that the BHT gravity  is the only  tree-level unitary model that can be constructed in 3D by judiciously combining  the Ricci scalar $R$ with  the  curvature-squared terms  $R^2$ and $R_{\mu \nu}^2$.
  \item To explore some interesting properties of this remarkable  model that have no counterpart in  the  usual Einstein gravity in three dimensional spacetime.  

\end{enumerate}
    
We describe in the following the steps we shall take in order to accomplish  these  objectives.  We start off our analysis by considering in section 2 the most general three-dimensional theory obtained by augmenting planar gravity through the  curvature-squared terms. Now, taking into account that in three dimensions both the curvature tensor and the Ricci tensor have the same number of components \cite{20}, we come to the conclusion that the Lagrangian density for the theory at hand can be written as 

\begin{eqnarray}
{\cal{L}} = \sqrt{g}\left(\frac{2 \sigma}{\kappa^2}R + \frac{\alpha}{2}R^2 + \frac{\beta}{2}R_{\mu \nu}^2  \right),
\end{eqnarray} 

\noindent where $\sigma$ is a convenient parameter that can take the values +1 (EH term with the standard sign), -1 (EH term with the ``wrong sign''),   and $\alpha$ and $\beta$ are free coefficients. Note that the constants $\kappa, \; \alpha,$ and $\beta$, have the mass dimension $[\kappa]= -\frac{1}{2}$
and $[\alpha]= [\beta]= -1$,  in fundamental units. We prove afterward that the BHT model is the only unitary system  at the tree level that can be built from the Lagrangian given in Eq. (2). In section 3 it is shown that, unlike what occurs in 3D general relativity, clocks are slowed down in a gravitational field described by the BHT model. This gravitational time dilation is the basis of the gravitational spectral shift. An expression for a new-massive-gravity-induced time delay is obtained in section 4. Finally, we present in section 5 some comments and observations.

We employ natural units, $c=\hbar=1$, and our Minkowski metric is diag(+1,\;-1,\; -1). Our Ricci tensor is defined by $R_{\mu \nu}= {R^\lambda}_{\mu \nu \lambda}\equiv \partial_\nu {\Gamma^\lambda}_{\mu \lambda} - \partial_\lambda{\Gamma^\lambda}_{\mu \nu} + ...$ . A  prescription for computing the graviton propagator, as well as  a list of some identities that greatly facilitate this task, are collected in Appendix A. The derivation of an important result for checking the  tree-level unitarity of a generic 3D gravity model is sketched in Appendix B.    

\section{Finding a class of tree-level unitary massive 3D gravity models}

To probe the unitarity at the tree level of the models defined by Eq. (2), we make use of an uncomplicated and easily handling algorithm that converts the task of checking the unitarity, which is in general a time-consuming work, into a straightforward algebraic exercise. The prescription consists basically in saturating the propagator with external conserved currents, compatible with the symmetries of the system, and in examining afterwards the residues of the saturated propagator (SP) at each simple pole.
Let us then compute the propagator for the gravity model in Eq. (2). To do that, we recall that for small fluctuations around the Minkowski metric $\eta$, the full metric assumes the form

\begin{equation}
g_{\mu\nu}=\eta_{\mu\nu}+\kappa h_{\mu\nu}
\end{equation}

Linearizing Eq. (2) via Eq. (3) and adding to the result the gauge-fixing Lagrangian density, $\mathcal{L}_{gf}=\frac{1}{2\Lambda}(\partial_\mu \gamma^{\mu\nu})^2$, where $\gamma_{\mu\nu}\equiv h_{\mu\nu}-\frac{1}{2}\eta_{\mu\nu}h$, that corresponds to the de Donder gauge, we find

\begin{equation}
\mathcal{L}=\frac{1}{2}h_{\mu\nu}\mathcal{O}^{\mu\nu,\alpha\beta}h_{\alpha\beta},
\end{equation}

\noindent where, in momentum space, 

\begin{eqnarray}
\nonumber\mathcal{O}&=&\left[\sigma k^{2}+\frac{\beta\kappa^{2}k^{4}}{4}\right]P^{(2)}+\frac{k^{2}}{2\Lambda}P^{(1)} +\frac{k^{2}}{4\Lambda}P^{(0-w)}-\frac{\sqrt{2}}{4}\frac{k^{2}}{\Lambda}P^{(0-sw)}\\&&-\frac{\sqrt{2}}{4}\frac{k^{2}}{\Lambda}P^{(0-ws)}
+\left[\frac{k^{2}}{2\Lambda}-\sigma k^{2}+2\alpha\kappa^{2} k^{4}+\frac{3}{4}\beta\kappa^{2} k^{4}\right]P^{(0-s)}.
\end{eqnarray}

\noindent Here $P^{(2)}$, $P^{(1)}$, $P^{(0-w)}$, $P^{(0-s)}$, $P^{(0-sw)}$ and $P^{(0-ws)}$ are the usual three-dimensional Barnes-Rivers operators (see Appendix A).

Therefore, the propagator is given by (see Appendix A)

\begin{eqnarray}
\nonumber\mathcal{O}^{-1}&=&\frac{2\Lambda}{k^2}P^{(1)}+\frac{1}{k^2(\sigma+\frac{\beta\kappa^2k^2}{4})}P^{(2)}+\frac{1}{-\sigma k^2+2\alpha\kappa^2k^4+\frac{3}{4}\kappa^2k^4\beta}P^{(0-s)} \\ \nonumber
&&+\frac{\sqrt{2}}{-\sigma k^2+2\alpha\kappa^2k^4+\frac{3}{4}\kappa^2k^4\beta}[P^{(0-sw)}+P^{(0-ws)}]\\&&+\frac{-4\Lambda\sigma+2+8\Lambda\alpha\kappa^2k^2+3\Lambda\beta\kappa^2k^2}{-\sigma k^2+2\alpha\kappa^2k^4+\frac{3}{4}\kappa^2k^4\beta}P^{(0-w)}.
\end{eqnarray}

Contracting now the above propagator with conserved currents $T^{\mu\nu}(k)$, ($k_{\mu}T^{\mu\nu}=k_{\nu}T^{\mu\nu}=0$), yields

\begin{eqnarray} {\mathrm{SP}}&=&\frac{1}{\sigma}\left[\frac{1}{k^{2}}-\frac{1}{k^{2}-m^{2}_{2}}\right]\left[T_{\mu\nu}^{2}-\frac{1}{2}T^{2}\right]+ \frac{1}{\sigma}\left[-\frac{1}{k^2} +\frac{1}{k^{2}-m_{0}^{2}}\right]\frac{1}{2}T^{2},
\end{eqnarray}

\noindent where $m^2_2\equiv-\frac{4\sigma}{\beta\kappa^2}$, $m_0^2\equiv\frac{4\sigma}{(8\alpha+3\beta)\kappa^2}$.
Assuming that there are no tachyons in the model, we promptly find the following constraints

\begin{equation}
\frac{\sigma}{\beta}<0, \qquad \frac{\sigma}{8\alpha+3\beta}>0.
\end{equation}

On the other hand, the residues of SP at the poles $k^2=m^2_2$, $k^2=0$, and $k^2=m^2_0$ are, respectively,

\begin{eqnarray}
Res ({\mathrm{SP}})\left. \right|_{k^{2}=m^{2}_{2}}&=&-\frac{1}{\sigma}\Big(T^{2}_{\mu\nu}-\frac{1}{2}T^{2}\Big)\left.\right|_{k^{2}=m_{2}^{2}},\\
Res({\mathrm{SP}}) \left.\right|_{k^2=0}&=&\frac{1}{\sigma}\Big(T^2_{\mu\nu}-T^2\Big)\left.\right|_{k^2=0},\\
Res({\mathrm{ SP}})\left.\right|_{k^2=m^2_0}&=&\frac{1}{2 \sigma}(T^2)\left.\right|_{k^2=m_0^2}.
\end{eqnarray}

Now, as is well-known, the  tree-level unitarity of a generic model is assured if the residue at each simple pole of ${\mathrm{SP}}$ is $\geq 0$. Keeping in mind that $\Big(T^2_{\mu\nu}-\frac{1}{2}T^2\Big)\left.\right|_{k^2=m_2^2}>0$ and $\Big(T^2_{\mu\nu}-T^2\Big)\left.\right|_{k^2=0}=0$ (see Appendix B), we arrive at the conclusion that: (i)$ Res ({\mathrm{SP}})\left.\right|_{k^{2}=m^{2}_{2}}>0$ if $\sigma=-1$ (which implies  $\beta>0$ and $\alpha<0$), and (ii) $Res ({\mathrm{ SP}})\left.\right|_{k^{2}=0}=0$. Consequently, we need not worry about these poles; the troublesome one is $k^2=m_{0}^{2}$ since $Res({\mathrm{ SP}})\left.\right|_{k^2=m_{0}^{2}}<0$. A way out of this difficult it is to consider the $m_{0}\rightarrow\infty$ limit of the model under discussion, which leads us to conclude that $\alpha=-\frac{3}{8}\beta$. Accordingly, the class of models defined by the Lagrangian density 

\begin{equation}
\mathcal{L}=\sqrt{g}\left[-\frac{2R}{\kappa^2}+\frac{\beta}{2}\left(R^2_{\mu\nu}-\frac{3}{8}R^2\right)\right],
\end{equation}
are ghost-free at the tree level. For the sake of convenience, we replace $\beta$ with $\frac{4}{\kappa^{2} m_2^{2}}$, where $m_2$ is a mass parameter. The resulting Lagrangian density, 

\begin{equation}
\mathcal{L}=\sqrt{g}\left[-\frac{2R}{\kappa^2}+\frac{2}{\kappa^2 m_2^2}\left(R^2_{\mu\nu}-\frac{3}{8}R^2\right)\right],
\end{equation}

\noindent is nothing but the BHT model for massive 3D gravity.

It is worth noting that it is not clear at all whether or not the particular ratio between $\alpha$ and $\beta$ we have previously found will survive renormalization at a given loop level, even at one-loop; in other words, unitarity beyond tree level has to be checked \cite{9}. Most likely  the BHT model is nonrenormalizable since it  improves only the spin-2 projections of the propagator but not the spin-0 projection \cite{21}.

\section{Gravitational time dilation}
Einstein 3D gravity is trivial outside the sources; consequently,  no  gravitational time dilation, or slowing down of clocks can take place in its framework. This can  easily be   shown  in the particular case of a spherically symmetric distribution of mass $M$ whose metric tensor is approximately given by 

\begin{eqnarray}
g_{\mu \nu} &=& \eta_{\mu \nu} + \kappa h_{\mu \nu} \nonumber \\ &=& 
\left(\begin{array}{ccc}
1&0&0\\
0&-(1+8GM\ln{\frac{r}{r_0}})&0\\
0&0&-(1+8GM\ln{\frac{r}{r_0}})	
\end{array}\right).
\end{eqnarray}

\noindent The corresponding spacetime interval  reads

\begin{eqnarray}
ds^2 =  dt^2 -(1 + \lambda)(dr^2 + r^2 d\theta^2),
\end{eqnarray}
\noindent where $\lambda = 8GM \ln \frac{r}{r_0}$, with $r_0$ being an infrared regulator, and $r$ and $\theta$ are the usual polar coordinates.

Introducing now new radial ($r'$) and angular ($\theta'$) coordinates through the change of variables

\begin{eqnarray}
(1-\lambda)r^2 = (1- 8GM)r'^2, \;\; \theta' = (1-4GM)\theta, \nonumber
\end{eqnarray}

\noindent we obtain, to linear order in $GM$,

\begin{equation}
ds^2 = dt^2 - dr'^2 - r'^2d\theta'^2. 
\end{equation} 

\noindent  The geometry around the spherically symmetric distribution  is, therefore,   locally identical to that of a flat spacetime as it should; however, it is not globally Minkowskian since the angle $\theta'$ varies in the range $ 0 \leq \theta' < 2\pi(1-4GM)$. Accordingly, the three-dimensional metric (16) describes a conical space with a wedge of angular size equal to $8\pi GM $ removed and the two faces of the wedge identified. We thus come to the conclusion that in the framework of Einstein 3D gravity no gravitational spectral-shift occurs due to   the presence  of the mentioned odd geometrical effect. It is worth noticing that in this context,  the non existence of  a time dilation does not imply that the spacetime is necessarily flat; in other words, the time dilation is not a ``classical test" of 3D general relativity. As we shall see in the following, the aforementioned bizarre  geometrical effect does not take place in new massive  gravity.  To do that we have to solve beforehand the linearized field equations related to the BHT system.

The field equations concerning the Lagrangian density

\begin{equation}
\mathcal{L}=\sqrt{g}\left[-\frac{2R}{\kappa^2}+\frac{2}{\kappa^2 m_2^2}\left(R^2_{\mu\nu}-\frac{3}{8}R^2\right) - \mathcal{L}_\mathrm{M}\right],
\end{equation}    

\noindent where  $\mathcal{L}_\mathrm{M}$ is the Lagrangian density for matter, are

\begin{eqnarray}
\nonumber&&G_{\mu\nu}+\frac{1}{m_{2}^{2}}\Bigg[\frac{1}{2}R^{2}_{\rho\sigma}g_{\mu\nu}-\frac{1}{4}\nabla_{\mu}\nabla_{\nu}R
-2R_{\mu\rho\lambda\nu}R^{\rho\lambda}
-\frac{1}{4}g_{\mu\nu}\Box R+\Box R_{\mu\nu}\\&&-\frac{3}{16}R^{2}g_{\mu\nu}+\frac{3}{4}RR_{\mu\nu}\Bigg]=\frac{\kappa^{2}}{4}T_{\mu\nu},
\end{eqnarray}
where $T_{\mu\nu}$ is the energy-momentum tensor, and $G_{\mu\nu}\equiv R_{\mu\nu}-\frac{1}{2}g_{\mu\nu}R$ is the Einstein tensor.

The corresponding linearized field equations are given by

\begin{equation}
\Bigg(1+\frac{\Box}{m_{2}^{2}}\Bigg)\Bigg[-\frac{1}{2}\Box h_{\mu\nu}+\frac{\eta_{\mu\nu}}{4\kappa}R^{\textrm{(lin)}}\Bigg]
+\frac{1}{2}\Big(\partial_{\mu}\Gamma_{\nu}+\partial_{\nu}\Gamma_{\mu}\Big)
=\frac{\kappa}{4}\Bigg(\frac{T}{2}\eta_{\mu\nu}-T_{\mu\nu}\Bigg),
\end{equation}
where $R^{\textrm{(lin)}}=\kappa \left[\frac{1}{2}\Box h-\gamma^{\mu\nu}_{\phantom{ab},\mu\nu}\right]$, $\Gamma_{\mu}\equiv\Big(1+\frac{\Box}{m_{2}^{2}}\Big)\partial_{\rho}\gamma_{\mu}^{\phantom{a}\rho}+\frac{\partial_{\mu}R^{(\textrm{lin})}}{4\kappa m_{2}^{2}}$, $\gamma_{\mu\nu}\equiv h_{\mu\nu}-\frac{1}{2}\eta_{\mu\nu}$. Note that here indices are raised (lowered) using $\eta^{\mu\nu}$($\eta_{\mu\nu}$).

Mimicking Teyssandier's work on 4D higher-derivative gravity \cite{22}, it can be shown that it is always possible to choose a coordinate system such that the gauge conditions, $\Gamma_\mu =0$, on the linearized metric, hold. Assuming that these conditions are satisfied, it is straightforward to show that the general solution of (19) is given by

\begin{equation}
h_{\mu \nu} = \psi_{\mu \nu} - h_{\mu \nu}^{(\mathrm{E})},
\end{equation}

\noindent where $h_{\mu \nu}^{(\mathrm{E})}$ is the solution of the linearized Einstein equation in the de Donder gauge, i.e.,

\begin{eqnarray}
\Box h_{\mu \nu}^{(\mathrm{E})} = \frac{\kappa}{2}(Tn_{\mu \nu} - T_{\mu \nu}), \; \; \partial^\nu \gamma_{\mu \nu}^{(\mathrm{E})}=0,   
\end{eqnarray}

 \noindent where $\gamma_{\mu \nu}^{(\mathrm{E})}\equiv h_{\mu \nu}^{(\mathrm{E})} - \frac{1}{2}\eta_{\mu \nu}h^{(\mathrm{E})}$, while $\psi_{\mu \nu}$ satisfies the equation 

\begin{equation}
(\Box + m^2_2)\psi_{\mu \nu}= - \frac{\kappa}{2}(T_{\mu \nu}- \frac{1}{2}\eta_{\mu \nu}T). 
\end{equation}

\noindent It is worth noticing that in this very special gauge the equations for $\psi_{\mu \nu}$ and $h_{\mu \nu}^{(\mathrm{E})}$ are totally decoupled. As a result, the general solution to the equation (19) reduces to a linear combination of  the solutions of the aforementioned equations.

Solving Eqs. (21) and (22) for a point-like particle of mass $M$ located at ${\bf r}= {\bf 0}$, we find

\begin{eqnarray}
h_{00}&=&-\frac{\kappa M}{8\pi}K_{0}(m_2r)\\
h_{11}&=&h_{22}=-\frac{\kappa M}{8\pi}\Big[K_{0}(m_2r)+2\ln\frac{r}{r_{0}}\Big],
\end{eqnarray}

\noindent where $K_0$ is the modified Bessel function of order zero. Note that $K_0(x)$ behaves as -ln$x$ at the origin and as $x^{-\frac{1}{2}}e^{-x}$
asymptotically. Thence, the metric tensor and the spacetime interval are given, respectively, by

\begin{eqnarray}\scriptsize{
g_{\mu \nu} = 
\left(\begin{array}{ccc}
1 -4MG K_0(m_2 r)&0&0\\
0&\begin{array}{l}-[1+4GM(K_0(m_2 r)\\\quad+ 2\ln{\frac{r}{r_0}})]\end{array}&0\\
0&0&\begin{array}{l}-[1+4GM(K_0(m_2 r)\\\quad+ 2\ln{\frac{r}{r_0}})]\end{array}

\end{array}\right),}
\end{eqnarray}

\begin{eqnarray}
ds^2 &=&[1 -4MGK_0(m_2 r)]  dt^2 -\left[1 + 4GM \left(K_0(m_2 r) + 2\ln\frac{r}{r_0}\right)\right](dr^2 \nonumber \\ &&+ r^2 d\theta^2).
\end{eqnarray}

\noindent In the $m_2 \rightarrow \infty $ limit, (25) and (26) reproduce (14) and (15), in this order, as expected. The geometry around the point-particle is, of course,  not locally identical to that of a Minkwoskian spacetime, signaling in this way the possibility of occurrence of gravitational spectral shift. Let us then show that the gravitational time dilation does occur in the BHT model.

Suppose that a signal sent from an emitter at a fixed point ($r_E, \theta_E$) is received, after traveling along a null geodesic,   by a receiver at a fixed point ($r_R, \theta_R$) (see Fig. 1). Now, the difference $t_R - t_E$, where $t_E$ is the coordinate time of emission and $t_R$ the coordinate time of reception, is the same for all signs sent --- the wordlines of successive signals are nothing but copies of successive signals merely shifted in time. As a result, if the $t$-time difference between a signal and the next is $dt_E$ at the departure point, the corresponding  $t$-time difference at the the arrival point is necessarily the same. However, the clock of an observer situated at the point of emission records  proper time ($\tau$) and not coordinate time ($t$). Accordingly, $d\tau_E= \sqrt{1- 4MGK_0(m_2 r_E)} dt_E$, and similarly  $d\tau_R= \sqrt{1- 4MGK_0(m_2 r_R)}dt_R$. Since $dt_E= dt_R$, we promptly obtain

\begin{eqnarray}
\frac{d\tau_R}{d\tau_E} &=&\frac{ \sqrt{1- 4MGK_0(m_2 r_R)}}{\sqrt{1- 4MGK_0(m_2 r_E)}} \nonumber \\ &\approx& 1- 2MGK_0(m_2 r_R) + 2MGK_0(m_2 r_E) \nonumber \\ &=& 1 + V_R - V_E, 
\end{eqnarray}

\noindent where $V(r) \equiv \frac{\kappa}{2} h_{00}(r)=-2MGK_0(m_2r)$ is the gravitational potential. This shows that if the clock at $(r_R,\theta_R)$ is at a lower potential than the clock at $(r_E,\theta_E)$, i.e., $V_R <V_E$, then $d\tau_R$ is smaller than $d\tau_E$. In other words, the clock that is deeper in the gravitational potential runs slower. Eq. (27) is the gravitational time-dilation formula, or redshift formula. It is worth noticing that $d\tau_R \rightarrow   d\tau_E$ as  $m_2 \rightarrow \infty$,  implying that no   gravitational time dilation takes place in the framework of 3D general relativity, which totally agrees with the result we have  previously found.

On the other hand, if  the emitter is a pulsating atom which in the proper time interval $\Delta \tau_E$ emits $n$ pulses, an observer situated at the emitter will assign to the atom a frequency $\nu_E \equiv \frac{n}{\Delta \tau_E}$, which, of course, is the proper frequency of the pulsating atom. The observer located at the receiver, in turn, assigns a frequency $\nu_R \equiv \frac{n}{\Delta \tau_R}$ to the pulsating atom. Consequently,

\begin{eqnarray}
\frac{\nu_R}{\nu_E} &=&\frac{ \sqrt{1- 4MGK_0(m_2 r_E)}}{\sqrt{1- 4MGK_0(m_2 r_R)}} \nonumber \\ &\approx& 1+  2MG\left[K_0(m_2 r_R) - K_0(m_2 r_E)\right]. \nonumber 
\end{eqnarray}

\noindent From this we immediately get the fractional shift

$$ \frac{\Delta \nu}{\nu} \equiv \frac{\nu_R - \nu_E}{\nu_E} \approx 2MG \left[ K_0(m_2 r_R) - K_0(m_2 r_E)\right].$$

\noindent Note that since $K_0(x)$ is a monotonically decreasing function in the range $0 \leq x < \infty, \;\frac{\Delta \nu}{\nu} $ is positive if $r_E > r_R$, and negative if $r_E < r_R$. Consequently, if the emitter is nearer to the massive object than the receiver is, the shift is towards the red, but if the receiver is nearer  the massive object, it is towards the blue. 

From the preceding considerations we come to the conclusion that the gravitational spectral shift is indeed a classical test of the BHT model. It can also be viewed, like in 4D general relativity,  as   a direct test of the curvature of the spacetime.

\begin{figure}[h]
\begin{center}
\includegraphics[scale=0.36]{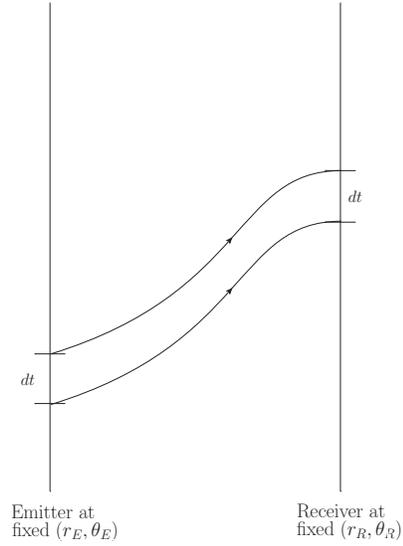}
\end{center}
\caption{\small Spacetime diagram illustrating the worldlines of two successive identical signals. }
\end{figure}
\section{Gravitational time delay}

Another interesting effect that can be obtained from the linear approximation of new massive gravity is the time delay suffered by a light signal sent by an observer --- situated at a fixed point in space in the gravitational field generated by a massive object --- to a small object and reflected back to the observer. The small object is supposed to be located directly between the observer and the huge body (see Fig. 2). Consider, in this spirit, a light pulse  that moves along a straight line connecting the observer and the small object. It is easy to show that the coordinate time for the whole trip (observer $\rightarrow $ small object $\rightarrow$ observer) is given by

\begin{eqnarray} 
\Delta t_G= 2\int_{r_1}^{r_2} \sqrt{\frac{1 + 4MG[K_0(m_2 r) + 2\ln \frac{r}{r_0} ]}{1- 4MGK_0(m_2 r)}}dr.
\end{eqnarray}

Accordingly, the proper time lapse measured by the observer, whose clock, of course, records proper time, has the form

\begin{eqnarray} 
\Delta \tau_G= 2 \sqrt{1- 4MGK_0(m_2 r_2)}	{\int_{r_1}^{r_2}}{\sqrt{\frac{1 + 4MG\left[K_0(m_2 r) + 2\ln \frac{r}{r_0}\right]}{1- 4MGK_0(m_2 r)}}dr.}
\end{eqnarray}

\begin{figure}[h]
\begin{center}
\includegraphics[scale=0.6]{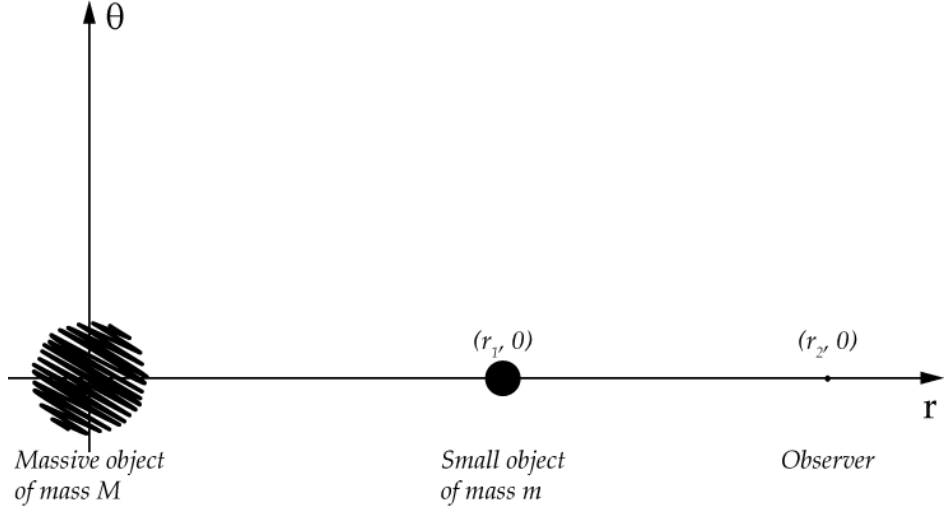}
\end{center}
\caption{\small Time delay in ``radar sounding". }
\end{figure}
 
On the other hand, the distance traveled by the light pulse is equal to $$2\int_{r_1}^{r_2} \sqrt{1 + 4MG[K_0 (m_2 r) + 2 \ln \frac{r}{r_0}]}dr.$$ Consequently, on the basis of the classical theory we should expect a round-trip time of 

\begin{eqnarray}
\Delta \tau_C = 2\int _{r_1}^{r_2}\sqrt{1+ 4MG\left[K_0(m_2 r) + 2\ln \frac{r}{r_0} \right]} dr.
\end{eqnarray}
 
  From (29) and (30), we arrive to the conclusion that $\Delta \tau_G \neq \Delta \tau_C$. Note that in the $m_2 \rightarrow \infty$ limit,   $\Delta \tau_G = \Delta\tau_C = 2 \int_{r_1}^{r_2} \sqrt{1 + 8MG \ln \frac{r}{r_0}} dr$, which clearly shows that there is no time delay in the framework of 3D general relativity, as expected.
 
 On the other hand, Eqs. (29) and (30), tell us that
 
 \begin{eqnarray}
 \Delta \tau_G &\approx& 2\int_{r_1}^{r_2}\left[ 1+ 4MG \left(K_0(m_2 r) +  \ln\frac{r}{r_0} \right) \right]dr \nonumber \\ &&- 4MG[K_0(m_2 r_2)](r_2 - r_1), \nonumber 
\end{eqnarray}
 
 \begin{eqnarray}
 \Delta \tau_C \approx 2\int_{r_1}^{r_2}\left[1 + 2GM\left(K_0(m_2 r) + 2\ln \frac{r}{r_0} \right)\right]dr. \nonumber
 \end{eqnarray}

As a result,

\begin{eqnarray}
\Delta \tau_G - \Delta \tau_C &\approx& 4MG\left[\int_{r_1}^{r_2}K_0(m_2 r)dr - (r_2 - r_1) K_0(m_2 r_2)\right] \nonumber \\ &=& 4MG \left[ K_0(m_2 r_0) - K_0(m_2 r_2)\right] (r_2 - r_1), \nonumber
\end{eqnarray}

\noindent where $r_1 < r_0 < r_2$. Hence, we come to the conclusion that there is a new-massive-gravity-induced time delay

\begin{eqnarray}
\Delta \tau_G - \Delta \tau_C \approx 4MG \left[K_0(m_2r_0) -K_0(m_2 r_2)\right] (r_2 - r_1).
\end{eqnarray}

\section{Final remarks}
As is well-known, three-dimensional Einstein gravity without sources is physically vacuous because Einstein and Riemann tensors are equivalent in $D=3$. In addition, the quantization of the gravity field does not give rise to propagating gravitons since the spacetime metric is locally determined by the sources. Consequently, the description of gravitational phenomena via $3D$ gravity leads to some bizarre results, such as the following.

\begin{itemize}
	\item Lack of a gravity force in the nonrelativistic limit.
	
	\item Gravitational deflection independent of the impact parameter. 
\item Complete absence of  gravitational time dilation.

\item No time delay.

\end{itemize}
 
It can be shown that the  first two  odd phenomena in the above list do not take place in the context of the BHT model \cite{23}. In fact, in the framework of the latter, short-range gravitational forces are  exerted on slowly moving particles; besides, the light bending depends on the impact parameter, as it should. On the other hand, the remaining strange phenomena in the aforementioned list, as we have shown,  do not occur in the BHT system either. Indeed, both time delay and spectral shift do take place   in the context of the new massive gravity. Like in 4D general relativity,  gravitational time dilation and gravitational time delay are also tests of the BHT model. It is worth noticing that the basis for these tests is the time-independent solution of the  linearized BHT field equations produced by  a static spherical mass.

One of the  main reasons for studying 3D gravity models is in reality to try to find out a gravity system with less austere ultraviolet divergences in perturbation theory. Since general relativity in 3D  is dynamically trivial, the BHT model, which is tree-level unitary, is an  important step in this direction.
 This kind of research conducted in lower dimensions certainly helps us to gain insight into difficult conceptional issues, which are present and more opaque in the physical (3+1)-dimensional world.  Another strong argument in favor of considering massive gravity theories, as we have already commented, is the fact that the present accelerated expansion of the universe could be partially attributed to a graviton mass-like effect.

It is worth mentioning that the triviality of 3D general relativity can also be cured by adding to the EH action in 3D a parity-violating Chern-Simons term. The resulting model is usually known as topological massive gravity (TMG) \cite{24,25}.  Nonetheless,  in contrast with TMG, 3D massive gravity has the great advantage of being  a parity-preserving theory. On the other hand, since 3D higher-derivative gravity (3DHDG)  --- which is defined by the Lagrangian density ${\cal{L}}_\mathrm{3DHDG} = \sqrt{g}\left(\frac{2\sigma}{\kappa^2} R + \frac{\beta}{2} R_{\mu \nu}^2 + \frac{\alpha}{2} R^2  \right)$ ---  is nonunitary at the tree level \cite{26}, it would be interesting to verify whether the addition of a topological Chern-Simons term (${\cal{L}}_\mathrm{CS}= \frac{\mu}{2} \epsilon^{\lambda\mu \nu} \Gamma^\rho_{\phantom{a}\sigma \lambda}[\partial_\mu \Gamma^\sigma_{\phantom{a}\rho \nu} + \frac{2}{3} \Gamma^\sigma_{\phantom{a}\omega \mu} \Gamma^\omega_{\phantom{a}\nu \rho} ], $ where $\mu$ is an arbitrary parameter) to this higher-order model would cure the nonunitarity of the former.  It can be shown that in order to avoid ghosts and tachyons in the mixed theory (${\cal{L}} = {\cal{L}}_\mathrm{3DHDG} + {\cal{L}}_\mathrm{CS}$)  the following constraints on the parameters must hold\footnote{The massless excitation, like the massless excitation of 3D general relativity, is a not a dynamical degree of freedom, i.e., it is nonpropagating.} \cite{27}:       

\begin{eqnarray*}
\textrm{(spin-2 sector)} &:& \sigma < 0, \; \beta >0, \\ 
\textrm{(spin-0 sector)} &:& \sigma > 0, \; 3\beta + 8\alpha > 0.
\end{eqnarray*}

\noindent Therefore, for arbitrary values of the parameters, the model at hand is nonunitary at the tree level, which clearly shows that the topological Chen-Simons term is not a panacea for 3DHDG's unitarity problem. Nevertheless, if we prevent the spin-0 mode from propagating by choosing $3\beta + 8\alpha =0$, the resulting model is tree-level unitary. It is amazing that the above condition is exactly the same constraint  that appears in the BHT model ($m_0 \rightarrow \infty$  limit). We call attention to the fact that, contrary to popular belief,   the addition of a Chern-Simons term to a tree-level unitary model is not necessarily a guarantee that the resulting model will be tree-level unitary \cite{26}. For instance, the addition of a Chern-Simons term (${\cal{L}}_\mathrm{CS}$) to three-dimensional $R + \alpha R^2$ gravity (${\cal{L}}_\mathrm{R+\alpha R^2} = (-\frac{2R}{\kappa^2} + \frac{\alpha R^2}{2}) \sqrt{g}$), which is tree-level unitary, spoils the unitary of the latter \cite{26}. Therefore, in some  cases the coexistence between the topological Chern-Simons term and 3D higher-derivative gravity theories is conflicting.

To conclude we remark that recently the nonlinear classical dynamics of the BHT model was exhaustively investigated  by de Rham, Gabadadze, Pirtskhalava, Tolley and Yavin \cite{28}, who found that the theory passed remarkably nontrivial checks at the nonlinear level, such as the following.

\begin{itemize}

\item In the decoupling limit of the theory, the interactions of the helicity-0 modes are described by a single cubic term, the so-called cubic Galileon \cite{29}.

\item The conformal mode of the metric coincides with the helicity-0 mode in the decoupling limit.

\item The full theory does not lead to any extra degrees of freedom, which suggests that a 3D analog of the 4D Boulware-Deser ghost is not present in the BHT system.

\end{itemize}

\section*{Acnowledgments}
 The authors are very grateful to FAPERJ, CNPq, and  CAPES (Brazilian agencies) for financial support.\\

\appendix \section{Propagator}

In order to find the propagator related to the Lagrangian density in Eq. (1) it is very convenient to work in terms of the Barnes-Rivers operators in the space of symmetric rank-two tensors. The complete set of 3-dimensional operators in momentum space is \cite{30,31}

\begin{eqnarray}
P^{(2)}_{\mu\nu,\kappa\lambda}&=&\frac{1}{2}(\theta_{\mu\kappa}\theta_{\nu\lambda}+\theta_{\mu\lambda}\theta_{\nu\kappa}-\theta_{\mu\nu}\theta_{\kappa\lambda}),\label{p2}\\
\nonumber P^{(1)}_{\mu\nu,\kappa\lambda } &=&\frac{1}{2}(\theta_{\mu\kappa}\omega_{\nu\lambda}+\theta_{\mu\lambda}\omega_{\nu\kappa}+\theta_{\nu\lambda}\omega_{\mu\kappa}+\theta_{\nu\kappa}\omega_{\mu\lambda}),\label{p1}\\
P^{(0-s)}_{\mu\nu,\kappa\lambda}&=&\frac{1}{2}\theta_{\mu\nu}\theta_{\kappa\lambda},\label{ps}\\
P^{(0-w)}_{\mu\nu,\kappa\lambda}&=&\omega_{\mu\nu}\omega_{\kappa\lambda},\label{pw}\\
P^{(0-sw)}_{\mu\nu,\kappa\lambda}&=&\frac{1}{\sqrt{2}}\theta_{\mu\nu}\omega_{\kappa\lambda},\label{psw}\\
P^{(0-ws)}_{\mu\nu,\kappa\lambda}&=&\frac{1}{\sqrt{2}}\omega_{\mu\nu}\theta_{\kappa\lambda}\label{pws},
\end{eqnarray}
where $\theta_{\mu\nu}\equiv\eta_{\mu\nu}-\frac{k_{\mu}k_{\nu}}{k^{2}}$ and $\omega_{\mu\nu}\equiv\frac{k_{\mu}k_{\nu}}{k^{2}}$ are, respectively, the usual transverse and longitudinal projection operators.
The multiplicative table for these operators is displayed in Table I.

\begin{table}[here]
	\centering
		\caption{Multiplicative table for the Barnes-Rivers operators}
		\begin{tabular}{l|cccccc}\\
			&$P^{(2)}$&$P^{(1)}$&$P^{(0-s)}$&$P^{(0-w)}$&$P^{(0-sw)}$&$P^{(0-ws)}$\\
			\hline
			$P^{(2)}$&$P^{(2)}$&0&0&0&0&0\\
			$P^{(1)}$&0&$P^{(1)}$&0&0&0&0\\
			$P^{(0-s)}$&0&0&$P^{(0-s)}$&0&$P^{(0-sw)}$&0\\
			$P^{(0-w)}$&0&0&0&$P^{(0-w)}$&0&$P^{(0-ws)}$\\
			$P^{(0-sw)}$&0&0&0&$P^{(0-sw)}$&0&$P^{(0-s)}$\\
			$P^{(0-ws)}$&0&0&$P^{(0-ws)}$&0&$P^{(0-w)}$&0
		\end{tabular}
	\label{BarnesRivers}
\end{table}

To compute the graviton propagator we need the bilinear part of the Lagrangian density (1). With the gauge fixing $\frac{1}{2\Lambda}(\partial_{\mu}\gamma^{\mu\nu})^2$ (de Donder gauge), and going over to momentum space we reproduce (5). The task of computing the operator $\mathcal{O}$ is greatly facilitated if we appeal to the following identities

\begin{eqnarray*}
&&\left[P^{(2)}+P^{(1)}+P^{(0-s)}+P^{(0-w)}\right]_{\mu\nu,\kappa\lambda}=\frac{1}{2}(\eta_{\mu\kappa}\eta_{\nu\lambda}+\eta_{\mu\lambda}\eta_{\nu\kappa}),\label{b1}
\nonumber\\
&&\left[2P^{(0-s)}+P^{(0-w)}+\sqrt{2}(P^{(0-sw)}+P^{(0-ws)})\right]_{\mu\nu,\kappa\lambda}=\eta_{\mu\nu}\eta_{\kappa\lambda},\label{b2}\\
\nonumber\\
&&\left[2P^{(1)}+4P^{(0-w)}\right]_{\mu\nu,\kappa\lambda}=\frac{1}{k^{2}}(\eta_{\mu\kappa}k_{\nu}k_{\lambda}+\eta_{\mu\lambda}k_{\nu}k_{\kappa}+\eta_{\nu\lambda}k_{\mu}k_{\kappa}+\eta_{\nu\kappa}k_{\mu}k_{\lambda}),\label{b3}\\
\nonumber\\
&&\left[2P^{(0-w)}+\sqrt{2}(P^{(0-sw)}+P^{(0-ws)})\right]_{\mu\nu,\kappa\lambda}=\frac{1}{k^{2}}(\eta_{\mu\nu}k_{\kappa}k_{\lambda}+\eta_{\kappa\lambda}k_{\mu}k_{\nu}),\label{b4}\\
\nonumber\\
&&P^{(0-w)}_{\mu\nu,\kappa\lambda}=\frac{1}{k^{4}}(k_{\mu}k_{\nu}k_{\kappa}k_{\lambda}).\label{b5}
\end{eqnarray*}

Now, if we write the operator $\mathcal{O}$ in the generic form
\begin{equation*}
\mathcal{O}=x_{1}P^{(1)}+ x_{2}P^{(2)}+x_{s}P^{(0-s)}+x_{w}P^{(0-w)}+x_{sw}P^{(0-sw)}+x_{ws}P^{(0-ws)},
\end{equation*}
and take into account that $\mathcal{O}\mathcal{O}^{-1}=I$, where $\mathcal{O}^{-1}$ is the propagator, we promptly find

\begin{eqnarray}
\nonumber\mathcal{O}^{-1}&=&\frac{1}{x_{1}}P^{(1)}+\frac{1}{x_{2}}P^{(2)}+\frac{1}{x_{s}x_{w}-x_{sw}x_{ws}}\Big[x_{w}P^{(0-s)}+x_{s}P^{(0-w)}\\&&-x_{sw}P^{(0-sw)}-x_{ws}P^{(0-ws)}\Big] .
\end{eqnarray}
  From (A.6) and (5) we obtain (6).

\section{A useful result}

\newtheorem{theorem}{Theorem}
\begin{theorem}
If $m$ is the mass of a generic physical particle related to a  given 3D gravitational model and $k$ is the corresponding exchanged momentum, then

\begin{equation*}
(T^2_{\mu\nu}-\frac{1}{2}T^2)|_{k^2=m^2}>0 \quad\mathrm{ and}\quad
(T^2_{\mu\nu}-T^2)|_{k^2=0}=0.
\end{equation*}
Here $T^{\mu\nu}(=T^{\nu\mu})$ is the external conserved current.
\end{theorem}

We begin by remarking that the set of independent vectors in momentum space, $k^{\mu}\equiv(k^{0}, \mathbf{k})$, $\tilde{k}^{\mu}\equiv(k^{0},-\mathbf{k})$, $\epsilon\equiv(0,\hat{\varepsilon})$, where $\hat{\varepsilon}$ is a unit vector orthogonal to $\mathbf{k}$, is  a suitable basis for expanding any three-vector  $V^{\mu}(k)$. Using this basis we can write the symmetric current tensor as follows
\begin{equation*}
T^{\mu\nu}=Ak^{\mu}k^{\nu}+B\tilde{k}^{\mu}\tilde{k}^{\nu}+C\epsilon^{\mu}\epsilon^{\nu}+Dk^{(\mu} \tilde{k}^{\nu)}+Ek^{(\mu}\epsilon^{\nu)}+F\tilde{k}^{(\mu}\epsilon^{\nu)},
\end{equation*}
where $a^{(\mu}b^{\nu)}\equiv\frac{1}{2}(a^{\mu}b^{\nu}+b^{\mu}a^{\nu})$.

The current conservations gives the following constraints on the coefficients $A$, $B$, $C$, $D$, $E$, and $F$:
\begin{eqnarray}
Ak^{2}+\frac{D}{2}(k_{0}^{2}+\mathbf{k}^{2})&=&0\\
B(k_{0}^{2}+\mathbf{k}^{2})+\frac{D}{2}k^{2}&=&0\\
Ek^{2}+F(k_{0}^{2}+\mathbf{k}^{2})&=&0
\end{eqnarray}

From Eqs. (B1) and (B2), we get $Ak^{4}+B(k_{0}^{2}+\mathbf{k}^{2})^{2}$, while Eq. (B3) implies $E^{2}>F^{2}$. On the other hand, saturating the indices of $T^{\mu\nu}$ with momenta $k_{\mu}$, we arrive at a consistent relation for the coefficients $A$, $B$, and $D$:

\begin{equation}
Ak^{4}+B(k_{0}^{2}+\mathbf{k}^{2})^{2}+Dk^{2}(k_{0}^{2}+\mathbf{k}^{2})=0.
\end{equation}

After a lengthy but otherwise straightforward calculation using the earlier equations, we obtain

\begin{eqnarray}
T_{\mu\nu}^{2}-\frac{1}{2}T^{2}&=&\Bigg[\frac{k^{2}(A-B)}{\sqrt{2}}-\frac{C}{\sqrt{2}}\Bigg]^{2}\nonumber +\frac{k^{2}}{2}(E^{2}-F^{2}),\\
T_{\mu\nu}^{2}-T^{2}&=&k^{2}\Bigg[\frac{1}{2}(E^{2}-F^{2})-2C(A-B)\Bigg].
\end{eqnarray}

Therefore,
$$
(T_{\mu\nu}^{2}-\frac{1}{2}T^{2})|_{k^2=m^{2}}>0 \quad\mathrm{and} \quad(T_{\mu\nu}^{2}-T^{2})|_{k^{2}=0}=0.
$$


\end{document}